\documentclass{nature}
\usepackage{graphicx}
\usepackage{amssymb}
\usepackage{dcolumn}
\usepackage{bm}
\usepackage{epstopdf}
\usepackage{epsfig}
\usepackage{color}
\usepackage{lineno}
\usepackage[colorlinks=true, letterpaper=true, pdfstartview=FitV, linkcolor=blue, citecolor=blue, urlcolor=blue]{hyperref}

\begin{document}

\title{Artificial gravity field, astrophysical analogues, and topological phase transitions in strained topological semimetals}

\author{Shan Guan$^{1,2}$, Zhi-Ming Yu$^{1,2}$, Ying Liu$^2$, Gui-Bin Liu$^1$, Liang Dong$^3$, Yunhao Lu$^4$, Yugui Yao$^{1,\star}$ \& Shengyuan A. Yang$^{2,\star}$}

\maketitle

\begin{affiliations}
\item Beijing Key Laboratory of Nanophotonics and Ultrafine Optoelectronic Systems, School of Physics, Beijing Institute of Technology, Beijing 100081, China
\item Research Laboratory for Quantum Materials, Singapore University of Technology and Design, Singapore 487372, Singapore
\item International Center for Quantum Materials, Peking University, Beijing 100871, China
\item School of Materials Science and Engineering, Zhejiang University, Hangzhou 310027, China

$^\star$e-mail: ygyao@bit.edu.cn;\, shengyuan\_yang@sutd.edu.sg
\end{affiliations}

\section*{Abstract}

\begin{abstract}
Effective gravity and gauge fields are emergent properties intrinsic for low-energy quasiparticles in topological semimetals. Here, taking two Dirac semimetals as examples, we demonstrate that applied lattice strain can generate warped spacetime, with fascinating analogues in astrophysics. Particularly, we study the possibility of simulating black-hole/white-hole event horizons and gravitational lensing effect. Furthermore, we discover strain-induced topological phase transitions, both in the bulk materials and in their thin films. Especially in thin films, the transition between the quantum spin Hall and the trivial insulating phases can be achieved by a small strain, naturally leading to the proposition of a novel piezo-topological transistor device. Possible experimental realizations and analogue of Hawking radiation effect are discussed. Our result bridges multiple disciplines, revealing topological semimetals as a unique table-top platform for exploring interesting phenomena in astrophysics and general relativity; it also suggests realistic materials and methods to achieve controlled topological phase transitions with great potential for device applications.
\end{abstract}

\noindent
{Keywords: Topological semimetal, Strain effects, Artificial gravity field, Topological phase transitions, Topological  transistor device}

\section*{Introduction}
Relativity is a fundamental aspect for all elementary particles in the high energy regime. In condensed matter physics, however, the relevant energy scale we probe is much lower (compared with e.g. the electron rest mass), hence the electronic dynamics is usually considered as non-relativistic.
Nonetheless, due to interactions with lattice and between electrons themselves, the electron properties in crystalline solids are strongly renormalized, and the resulting low-energy electron quasiparticles can behave drastically different from free electrons. Remarkably,
in a class of recently discovered topological semimetal materials, the band structures feature nontrivial band-crossings close to the Fermi level, around which the low-energy quasiparticles become massless and resemble relativistic particles. For example, in so-called Weyl semimetals,
the Fermi surface consists of isolated band-crossing points, each carrying a topological charge of $\pm 1$ corresponding to its chirality, and the low-energy quasiparticles mimic the Weyl fermions in high-energy physics~\cite{Wan2011,Murakami2007}. With further protection from crystalline symmetry, a pair of Weyl points can be stabilized at the same point (called the Dirac point) in the energy-momentum space, realizing the Dirac semimetal (DSM) phase~\cite{Young2012}.
A number of 3D materials have been predicted to host Weyl/Dirac points~\cite{Burkov2011,Lu2013,Xu2014,Yang2014,Weng2015,Huang2015,Li2017}. Some of them, including the DSMs Na$_3$Bi and Cd$_3$As$_2$~\cite{Wang2012,Wang2013}, have been confirmed in recent experiments~\cite{Liu2014,Borisenko2014,Liu2014a,Jeon2014,Xu2015,Lv2015a,Shekhar2015,Yang2015}.

These topological semimetals offer a versatile platform for simulating relativistic particles and their many fascinating phenomena~\cite{Nielsen1983,Son2012,Lundgren2014}. Indeed, the nontrivial topology of the band-crossing point dictates the emergence of Lorentz symmetry for the low-energy quasiparticles~\cite{Volovik2003,Horava2005,Zhaoyx2013,Cortijo2016}: the inverse quasiparticle propagator can be written in a general and manifestly covariant form
\begin{equation}\label{Ginv}
\mathcal{G}^{-1}=\sigma^\alpha e^\mu_\alpha (p_\mu-p_\mu^{(0)}),
\end{equation}
where $p_\mu$ is the covariant energy-momentum four-vector, $p_\mu^{(0)}$ corresponds to the location of the Weyl point, $\sigma^\alpha=(1,\bm\sigma)$ with $\bm \sigma$ the vector of Pauli matrices, and all material-specific model parameters are encoded in the tetrad field $e^\mu_\alpha$. Consequently, the relativistic spectrum follows the equation:
\begin{equation}\label{Spec}
g^{\mu\nu}(p_\mu-p_\mu^{(0)})(p_\nu-p_\nu^{(0)})=0,
\end{equation}
where $g^{\mu\nu}=\eta^{\alpha\beta} e^\mu_\alpha e^\nu_\beta$ ($\eta^{\alpha\beta}=\textup{diag}(-1,1,1,1)$)
plays the role of an effective spacetime metric and $p_\mu^{(0)}$ acts as an effective $U(1)$ gauge potential, constituting the ``fermionic vacuum" where the low-energy quasiparticles live. Here $g^{\mu\nu}$ and $p_\mu^{(0)}$ are fully determined by the material band structure, and their spatial and temporal variation may give rise to effective gravity and gauge fields respectively. As a direct manifestation of the emergent relativistic symmetry, the effective spacetime metric $g^{\mu\nu}$ offers intriguing possibility to probe effects from general relativity in solid-state systems. Such possibility has not been explored so far.

Meanwhile, topological semimetals represent the neighbor state to various topological quantum phases. Particularly, DSMs are long believed to be an ideal platform for the systematic study of topological phase transitions (TPTs). However, a realistic approach to achieve controllable and reversible TPTs is still missing. Such an approach is much desired because of its significance in both fundamental physics investigation and in potential topological device applications.

In this work, we show that both objectives mentioned above can be achieved via lattice strain in topological semimetals. As a notable advantage, the solid-state systems admit an easy tuning of their properties by strain. Taking two DSMs as concrete examples and with first-principles calculations, we demonstrate that the quasiparticle spectrum can be efficiently controlled by uniaxial strain. We show that an inhomogeneous strain profile can generate warped spacetime with fascinating analogues in astrophysics. Particularly, we analyze the possibility to simulate black-hole/white-hole event horizons and gravitational lensing effect. Furthermore, we find that a larger strain can completely change the fermionic vacuum, leading to a TPT between DSM and trivial insulator phases, during which the two Dirac points collide at the $\Gamma$ point and pair-annihilate. More importantly, for a quasi-2D DSM thin film, a small strain is sufficient to control a TPT between quantum spin Hall (QSH) and trivial insulator phases. This is regarded as a key to the realization of a topological transistor. The discovery here allows us to propose a novel piezo-topological transistor device.
Thus our work not only establishes bridges between distinct disciplines such as condensed matter physics, astrophysics, and general relativity, it also reveals realistic platform and methods to study the intriguing topological phase transitions and to achieve unprecedented device functionalities for applications.

\section*{Results}

\subsection{Strain tuning and TPT in bulk material.}
In this work, we take the two DSMs Na$_3$Bi and Cd$_3$As$_2$ as concrete examples to demonstrate our general idea. These two materials represent the first two topological semimetals that have been confirmed by experiment~\cite{Liu2014,Borisenko2014,Liu2014a,Jeon2014}. We choose them as examples because they share a relatively simple low-energy bulk band structure. Both materials have a single pair of Dirac points located at $k_z=\pm k_D$ on the $k_z$-axis near the $\Gamma$ point of the Brillouin zone~\cite{Wang2012,Wang2013}. Each Dirac point is four-fold degenerate, consisting of two Weyl points of opposite chirality. The decoupling of the two Weyl points (hence the stability of the Dirac point) is protected by the $C_3$ ($C_4$) rotational symmetry of Na$_3$Bi (Cd$_3$As$_2$). The band ordering is inverted at the $\Gamma$ point, as it should be for the realization of band-crossing points.

In this work, we focus on the uniaxial strain along the crystalline $c$-axis ($z$-direction) (see Figure 1a), which preserves the corresponding rotational symmetry hence the existence of Dirac points. The low-energy effective model can be written for each Dirac point. For example, the quasiparticles around the Dirac point at $k_z=+k_D$ are described by two copies of the Weyl Hamiltonian~\cite{Wang2012,Wang2013}:
\begin{equation}\label{Heff}
\mathcal{H}_\pm=\pm v_\bot k_x\sigma_x +v_\bot k_y \sigma_y+v_z (k_z-k_D)\sigma_z
+w (k_z-k_D),
\end{equation}
each with a definite chirality corresponding to the subscript of $\mathcal{H}$. Here $v_\bot$ and $v_z$ are the Fermi velocities in the $xy$-plane and along the $z$-axis respectively (we set $\hbar=1$), and the last term in Eq.(\ref{Heff}) tilts the spectrum along $k_z$. The model for the other Dirac point at $k_z=-k_D$ can be simply obtained from (\ref{Heff}) by a time reversal operation.

Under moderate uniaxial strain, the form of model (\ref{Heff}) is preserved, only the model parameters change with strain. The parameters can be evaluated by fitting the first-principles band structure. Since the two materials show qualitatively similar behavior, in the following presentation, our discussion will be mainly based on the results of Na$_3$Bi. (The first-principles calculation method and the results for Cd$_3$As$_2$ are presented in the Supplementary Information.)
The band structures of Na$_3$Bi for several representative strains are shown in Figure 1b-e. Here strain $\varepsilon=\frac{\ell-\ell_0}{\ell_0}$ where $\ell$ is the lattice parameter along $c$-axis and $\ell_0$ is its equilibrium value.
The values of model parameters versus strain are plotted in Figure 2. Importantly, one observes that with compressive strain, $k_D$ decreases and approaches zero at a critical strain $\varepsilon_c\sim -6.2\%$. This means that the locations of the two Dirac points are shifted towards the $\Gamma$ point by strain and collide with each other at $\varepsilon_c$. Beyond this point, the effective model (\ref{Heff}) is no longer valid. From the first-principles band structure (Figure 1c-e), one can see that the two Dirac points annihilate with each other and eventually a finite gap is opened in the spectrum, leading to a band insulator at large strains. In this process, the direct gap at the $\Gamma$ point, which indicates the strength of band inversion, shrinks and changes sign at $\varepsilon_c$, marking a reversal of band ordering around the $\Gamma$ point (from inverted ordering to normal ordering). Thus, the transition at $\varepsilon_c$ represents a TPT from a DSM phase to a trivial insulator phase (the change in Fermi surface topology means that it is also a Lifshitz transition).

Before proceeding, we note that, first, the value of critical strain is correlated with the band inversion strength: Cd$_3$As$_2$ has a smaller inverted gap at the $\Gamma$ point than Na$_3$Bi, hence its critical strain ($\sim-1.3\%$) is also smaller, making the TPT comparatively easier to achieve. Second, the TPT completely changes the fermionic vacuum, from that of massless Dirac particles to massive particles with a finite excitation gap, which should be easily probed in spectroscopic or transport experiment.

\subsection{Artificial gravity field and astrophysical analogues.}
Now let's consider the regime of small strains, for which the system lies within the DSM phase away from the TPT, such that the quasiparticles are well-described by model (\ref{Heff}). Artificial fields are generated when we allow spacetime variation of the applied strain. Here we focus on static strain profiles and require that the strain is slowly-varying on the scale of lattice constant, i.e. $|\nabla \varepsilon|\ll \ell_0^{-1}$, such that the strain effect can be captured by a local Hamiltonian with spatially-dependent parameters (as in Eq.(\ref{Heff})), forming a smooth background where the quasiparticles move around. Below, in discussing the quasiparticle propagation in the presence of a nontrivial spacetime metric, we adopt a quasi-classical description, which would further require that the strain is slowly-varying compared with the Fermi wavelength ($|\nabla \varepsilon|\ll \lambda_F^{-1}$).

The effective spacetime metric can be obtained by a direct comparison of model (\ref{Heff}) with Eqs.(\ref{Ginv}) and (\ref{Spec}). Then the coordinate differential, which characterizes the spacetime geometry, can be obtained as
\begin{equation}
ds^2=g_{\mu\nu}dx^\mu dx^\nu=-dt^2+\frac{1}{v_\bot^2}(dx^2+dy^2)+\frac{1}{v_z^2}(dz-wdt)^2.
\end{equation}
Interestingly, one observes that the tilt parameter $w$ mixes the space and time components. Its effect is like viewing a untilted spectrum (with $w=0$) in a moving reference frame with speed $w$. Indeed, in retrospect, one realizes that the tilt term in Eq.(\ref{Heff}) is just the Doppler shift when (Galilean) transformed to the moving frame. With inhomogeneous static strain, the parameters $v_\bot$, $v_z$ and $w$ become functions of spatial coordinates. In the following, we consider two simple example configurations.

In the first example, we focus on the quasiparticle motion along the $z$-direction, assuming parameters depend only on $z$ and ignoring the $x$ and $y$ coordinates. With a general coordinate transformation $\bar{t}=t+\int^z w(z')dz'/[v_z^2(z')-w^2(z')]$, the effective metric can be written as
\begin{equation}\label{1D}
ds^2=-\left[1-\left(\frac{w}{v_z}\right)^2\right]d\bar{t}^2+\frac{1}{v_z^2}\frac{dz^2}{\left[1-\left(\frac{w}{v_z}\right)^2\right]}.
\end{equation}
One notes that this metric shares the same form as the radial part of the familiar Schwarzschild metric for a spherical gravitating source~\cite{Volovik2003,Cheng2010}. From the analogy, one can directly obtain the effective gravitational potential $\Phi(z)=-\frac{1}{2}(\frac{w}{v_z})^2$ (here defined as dimensionless, as in units of $v_z^2(\infty)$) and the corresponding gravitational field $-\frac{d}{dz}\Phi \hat{z}$. The analogy can be made more precise if we design the strain profile such that $\Phi(z)\propto -\frac{1}{z}$ (which can be done in certain region excluding the $z=0$ singularity), simulating the gravity of an object located at $z=0$ with a mass of $-z\Phi(z)/G$, where $G$ is the Newton's constant.

The Schwarzschild metric has a coordinate singularity at the so-called Schwarzschild radius corresponding to an event horizon, where the space-like and time-like coordinates switch roles~\cite{Cheng2010}. One naturally speculates the possibility of similar physics here. In Eq.(\ref{1D}), this occurs where the value $|w/v_z|$ crosses 1. The underlying physics can be easily understood as in Figure 3a. Assuming that $v_z,w>0$, and $w/v_z>1(<1)$ for $z<z_h(>z_h)$ denoted as region A (B). Then from the quasiparticle spectrum, in region B, we have both left- and right-propagating modes; whereas in region A, since the tilt $w$ dominates over the Fermi velocity $v_z$, the spectrum is tipped over, as a result, only the right-propagating modes exist (see Figure 3a top panel). This means that any quasiparticle in region A must cross the point $z=z_h$ and be emitted into region B. Therefore this point represents a white-hole horizon for the quasiparticles. In addition, due to time reversal, the quasiparticles at the other Dirac point would observe $z=z_h$ as a black-hole horizon: a particle crosses the horizon from region B to region A cannot get back (see Figure 3a bottom panel).

For the two DSM materials considered here, we do not find the case with $|w/v_z|>1$ (at least for the uniaxial strain considered here). In fact, this case corresponds to so-called type-II Weyl/Dirac points, which has recently been predicted in several materials~\cite{Soluyanov2015,Xu2015b,Ruan2016,Chang2016}. Hence according to our analysis, an event horizon can in principle be realized at the boundary between type-I and type-II regions. It should be noted that for type-II materials, the Fermi surface geometry completely changes, typically involving multiple electron and hole pockets. The above discussion holds only for the quasiparticles close to the Weyl/Dirac point; for quasiparticples away from the point, they will not necessarily perceive the event horizons. Nevertheless, the horizon in such case still has physical meaning as the phase boundary separating two fermionic vacua with different topology of the spectrum: with Fermi point on one side and with Fermi surface on the other side.

In the second example, we focus on the quasiparticle motion in the $xy$-plane (ignoring the $z$ dimension) and consider the analogue of gravitational lensing effect. Then in terms of polar coordinates, we find that
\begin{equation}
ds^2=-\left[1-\left(\frac{w}{v_z}\right)^2\right]dt^2+\frac{1}{v_\bot^2}(d\rho^2+\rho^2 d\phi^2).
\end{equation}
Following the standard procedure~\cite{Cheng2010}, one finds the quasiparticle effective ``speed of light" viewed by a remote observer at $\rho=\infty$,
\begin{equation}
c(\bm \rho)=v_\bot\sqrt{1-\left(\frac{w}{v_z}\right)^2}.
\end{equation}
Then the propagation of the quasiparticle can be conveniently described by an effective vacuum index of refraction $n(\bm \rho)=c(\rho=\infty)/c(\bm\rho)$, naturally leading to the bending of particle trajectories (corresponding to the geodesic in the warped spacetime) with inhomogeneous strains, like in geometric optics. One notes that here the same $n$ holds for both Dirac points. The variation of $n$ versus strain is plotted in Figure 3b.

To simulate the real gravitational lensing in astronomy, one can design a strain profile (using Figure 3b), such that $n(\rho)=[1+2\Phi(\rho)]^{-1}$ with $\Phi(\rho) \propto -\frac{1}{\rho}$ in a region excluding the singularity at $\rho=0$. Assuming the quasiparticle trajectory lies in this region and the variation of $n$ is small and smooth, the classical gravitational lensing result directly applies~\cite{Cheng2010}: the deflection angle $\delta \phi$ of a quasiparticle trajectory (see Figure 3c) with the closest approaching distance $\rho_{min}$ ($\approx$ the impact parameter) is given by
\begin{equation}
\delta\phi\approx -4\Phi(\rho_{min}).
\end{equation}
Interestingly, since one can control both the vacuum state at $\rho=\infty$ and the sign of applied strain, it is possible to realize $n(\rho)<1$,  which corresponds to an anti-gravitating source, for which the quasiparticle trajectories are repelled from the source (Figure 3d).

\subsection{TPT in thin film and piezo-topological transistor.}
The TPT in the bulk typically requires a large strain (which is still within the linear elastic regime for the two materials, as shown in Supplementary Information). In the following, we show that a related TPT is more readily achievable in a DSM thin film by small strains.

First of all, one notes that in the bulk band structure of both DSMs, the gap is inverted in-between the two Dirac points, i.e., considering a 2D slice of the bulk Brillouin zone perpendicular to the $k_z$-axis, its gap is inverted if the slice lies in-between the two Dirac points, and is non-inverted otherwise. For a Na$_3$Bi (or Cd$_3$As$_2$) thin film confined in $z$-direction~\cite{Hellerstedt2016,Moll2016}, the electron motion along $z$ is quantized into discrete quantum well levels, forming quantum well subbands in the spectrum~\cite{Xiao2015,Pan2015}. The resulting system generally becomes semiconducting. In the quantum well approximation~\cite{Liu2010}, each subband corresponds to a 2D slice in the original 3D band structure with an effective wave-vector $(\tilde{k}_z)_m=m\pi/L$, where the integer $m(=1,2,\cdots)$ labels the subbands and $L$ is the film thickness. Thus the gap of the $m$th-subband is inverted (non-inverted) if $(\tilde{k}_z)_m<k_D(>k_D)$. Each inverted subband contributes a nontrivial 2D $\mathbb{Z}_2$-invariant~\cite{Shen2012}, so that the quasi-2D thin film becomes a QSH insulator if there is an odd number of inverted subbands, and is a trivial insulator if this number is even. This mechanism has been revealed in the topological phase oscillation versus $L$~\cite{Wang2013,Xiao2015}.

Now, under uniaxial strain, both $k_D$ and $(\tilde{k}_z)_m$ will change. Consider the case when the $m$th-subband has the smallest gap and has $(\tilde{k}_z)_m<k_D$. With compressive strain, $(\tilde{k}_z)_m$ increases, whereas $k_D$ decreases. So there must exist a critical strain $\tilde{\varepsilon}_c$ where $(\tilde{k}_z)_m$ and $k_D$ cross each other (see Figure 4a). In the process, the subband gap closes and re-opens with a switch of band ordering. This changes the $\mathbb{Z}_2$ character of the whole system by 1, leading to a TPT between a QSH insulator and a trivial insulator phases, as illustrated in Figure 4b. Since this TPT does not need $k_D$ to vanish as in the bulk case, the required strain can be much smaller (in the example shown in Figure 4, the critical strain is $\sim-1\%$).

The QSH state possesses topologically-protected gapless edge channels, in which the carriers can transport without back-scattering~\cite{Hasan2010,Qi2011}.
This leads to the proposition of so-called topological transistor based on QSH channel materials, which is expected to have the advantages of fast operating speed, low heat dissipation, and low power consumption~\cite{Qian2014}. So far, QSH state has been confirmed only in a few quantum well structures~\cite{Hasan2010,Qi2011}, and how to reliably control the TPT (hence the switch between on and off states) remains a challenge for designing topological transistor~\cite{Pan2015}. Our discovery here points to a promising novel device---a piezo-topological transistor. As illustrated in Figure 5, this device has a DSM thin film as the channel. Suppose the thin film without strain is in the trivial insulator phase, for which the transistor is at the off state (Figure 5a). By applying a small strain, the layer can be driven to the QSH phase, with current conduction through topological edge channels, corresponding to the on state (Figure 5b). Like previous proposals, this device enjoys advantages such as low dissipation and robust operation; besides, the sensitivity to strain makes it promising for electromechanical sensing applications.

\section*{Discussion}

The emergence of relativistic spectrum and Lorentz invariance at low energy is a general feature dictated by the Fermi point topology.  Here we take the two specific materials as examples, but the underlying physics is general and applies to other topological semimetals as well. Similar relativistic spectrum was previously discussed in the superfluid $^3$He-A phase, where analogues of black holes were suggested by controlling vortices and background superfluid flow~\cite{Unruh1981,Volovik2003,Volovik2016}. In a recent experiment, an artificial black hole for acoustic waves was simulated in an accelerating atomic Bose-Einstein condensate at sub-Kelvin temperature~\cite{Steinhauer2016}. In comparison, the solid-state system studied here can work at room temperature and permits much easier control by static means like strain or external fields, hence the predicted effects should be more readily observable.

Many experimental techniques have been developed for engineering strain, such as nanoindentation, micro-compression testing, and by using profiled substrate or stretchable substrate for thin film structures. Specifically, regarding the two examples with inhomogeneous strain profiles discussed here, the first one may be realized by using a setup similar to the micro-pillar compression test~\cite{Jiang2010}, and the strain variation along vertical direction could be achieved either by engineering the geometric shape of the pillar or by applying profiled lateral constriction to the side of the pillar. For the second example, an $xy$-dependent strain profile may be achieved by a setup similar to nanoindentation, e.g. by using a hard tip with engineered shape to press onto the Dirac semimetal slab. We notice that this kind of setup has recently been utilized to study the pressure-induced superconductivity in Cd$_3$As$_2$~\cite{Wang2016a}. Furthermore, we point out that the lattice strain would generally be inhomogeneous in real strained samples, and more complicated strain profiles could be naturally realized. Given that the current technology can map out strain distribution with nanometer spatial resolution and with a precision $\sim0.1\%$~\cite{Hytch2008}, as long as the strain is slowly-varying, one can compute the effective metric and study the quasiparticle propagation under artificial gravity using the idea presented here.

We have mentioned that the spatial and/or temporal variation of the Weyl/Dirac point location will give rise to effective gauge fields. Here, $k_D$ in Eq.(\ref{Heff}) acts as the $z$-component of a vector potential, its spatial variation generally leads to a pseudo-magnetic field in the $xy$-plane. For example, for a strain varying along the $x$-direction, the induced pseudo-magnetic field is along the $y$-direction, with a magnitude given by
$B=\frac{1}{e}\frac{dk_D}{d\varepsilon}\frac{d\varepsilon}{dx}$. For a strain variation of $3\%$ over 10 nm, the field strength can be up to 16.5 T for Na$_3$Bi. It should be noted that the field direction is reversed for the other Dirac point, as required by the time reversal symmetry. Similar pseudo-magnetic field has been studied particularly in strained graphene~\cite{Levy2010}, and recently also discussed in topological semimetals~\cite{Cortijo2015,Pikulin2016}.

Our treatment of the artificial gravity here is within a quasi-classical approach, which leaves out possible analogues of interesting quantum mechanical effects in curved spacetime. For example, Hawking radiation~\cite{Hawking1974}, the radiation of particles from the black-hole horizon stimulated by quantum vacuum fluctuation, may find analogue effects here. Assuming the presence of a black-hole horizon in a topological semimetal setting (as in Figure 3a), the quantum tunneling process as illustrated in Figure 6 can be interpreted as the creation of a pair of fermions due to quantum fluctuation: a quasiparticle outside the horizon and a quasihole inside the horizon. The quasiparticle escapes from the horizon, resembling the Hawking radiation. With the metric in Eq.(\ref{1D}) and by direct analogy~\cite{Hawking1974,Volovik2003}, one can find the associated Hawking temperature $T_\mathrm{H}=\frac{\hbar}{2\pi k_B}|\frac{d}{dz}(v_z-w)|_{z_h}$ for the radiation. Nevertheless, an accurate study of these quantum effects and the methods to probe them is beyond the scope of this work and deserves a future investigation.

Finally, we note that a similar semimetal-to-insulator TPT was predicted in bulk Na$_3$Bi$_{1-x}$Sb$_x$ and Cd$_3$[As$_{1-x}$P$_x$]$_2$ alloys by tuning the substitute concentrations~\cite{Narayan2014}. In comparison, the strain approach here has the obvious advantage of allowing a reversible and continuous TPT to be achieved. For application purpose, the TPT in the thin film seems more appealing because of the dissipationless and spin-filtered transport associated with the QSH edge channels. By controlling the spatial strain profile, one can imagine to make a thin film with patterned QSH regions and trivial insulating regions, forming a topological circuit with designed 1D helical spin channels. It will provide an ideal platform for integrating various topological devices and functionalities to achieve unprecedented performance.


\begin{addendum}
\item [Acknowledgements]
The authors thank D.L. Deng for helpful discussions. This work was supported by the MOST Project of China (Nos 2014CB920903 and 2016YFA0300603), the National Natural Science Foundation of China (Grant Nos 11574029, 11225418, 11374009, 61574123, and 21373184), National Key Basic Research Program of China (2012CB825700), and Singapore MOE Academic Research Fund Tier 1 (SUTD-T1-2015004) and Tier 2 (MOE2015-T2-2-144).

\item [Author Contributions]
S.G., G.-B.L., and Y.H.L. performed the first-principles calculation and the data analysis.
Z.-M.Y., Y.L., L.D., and S.A.Y. performed the analytical modeling and calculation. Y.Y. and S.A.Y. supervised the work. All authors contributed to the discussion and reviewed the manuscript.

\item [Competing Interests]
The authors declare no competing financial interests.

\item [Correspondence]
Correspondence should be addressed to Yugui Yao or Shengyuan A. Yang.

\item [Additional Information]
Supplementary information (including the details of the first-principles method, the DFT results for Cd$_3$As$_2$, mechanical properties of the two DSMs, and the low-energy effective model) is available in the online version of the paper.

\end{addendum}

\newpage
\begin{figure}
  \begin{center}\label{fig1}
   \epsfig{file=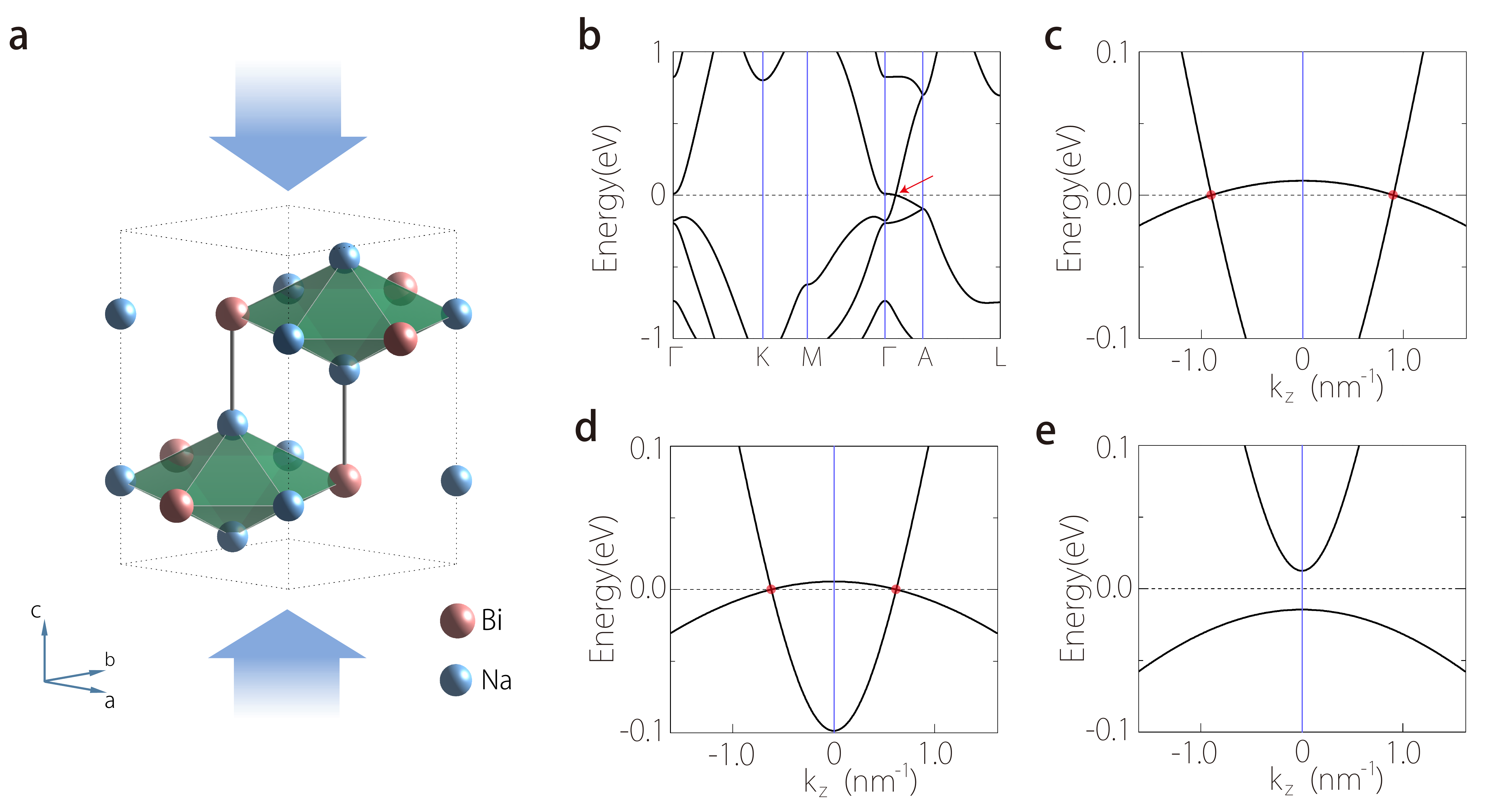,width=16cm}
  \end{center}
  \caption{\textbf{First-principles band structure results for Na$_3$Bi.} (\textbf{a}) Schematic representation of Na$_3$Bi lattice structure under uniaxial strain. (\textbf{b}) Band structure without strain. (\textbf{c-e}) Enlarged band structures near the $\Gamma$ point along the $k_z$-axis for strain value (c) $0\%$, (d) $-3\%$, and (e) $-7\%$. The Dirac points are indicated by the red arrow in (b) and by red dots in (c,d). }
\end{figure}
\newpage
\begin{figure}
  \begin{center}\label{fig2}
   \epsfig{file=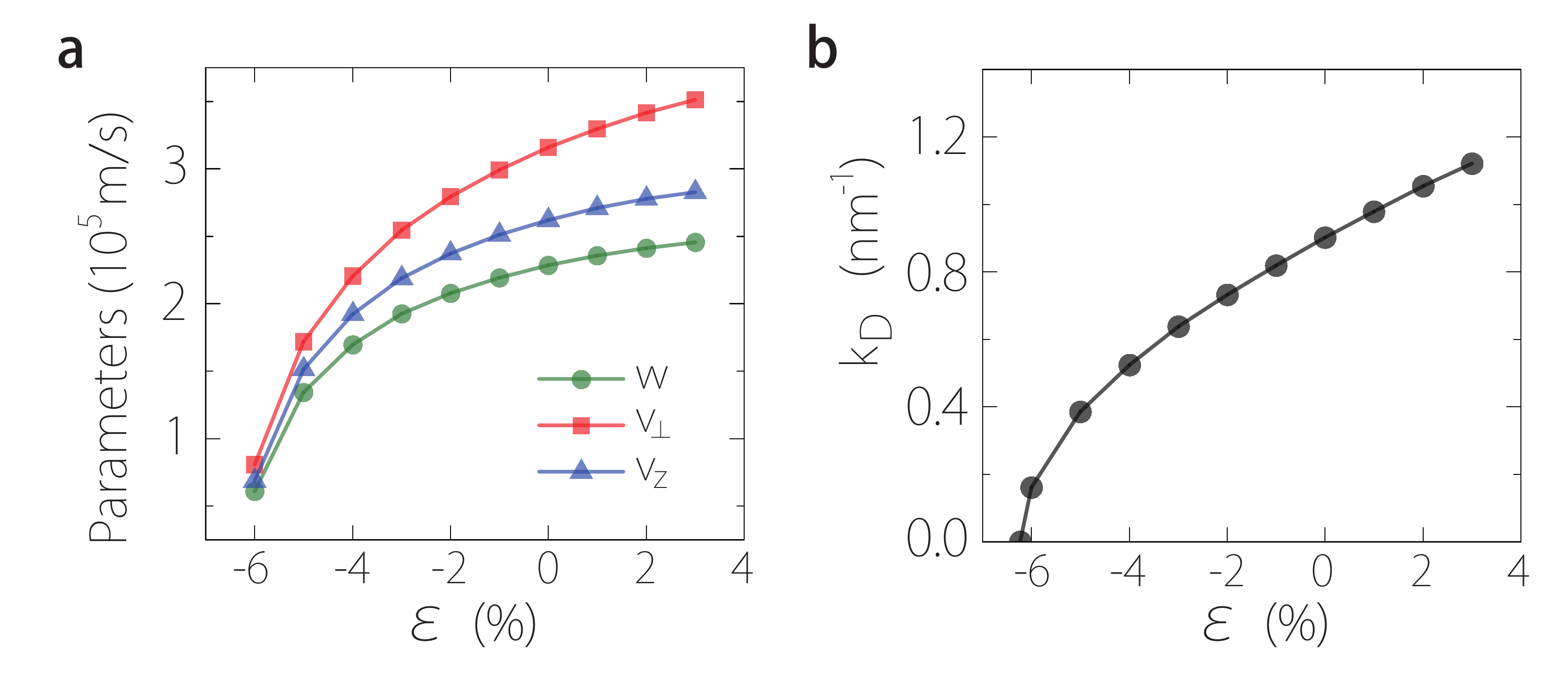,width=16cm}
  \end{center}
  \caption{\textbf{Model parameter versus strain.} The model parameters (\textbf{a}) $v_\bot$, $v_z$, $w$, and (\textbf{b}) $k_D$ in Eq.(\ref{Heff}) as a function of strain obtained by fitting the first-principles band structures of Na$_3$Bi.}
\end{figure}
\newpage
\begin{figure}
  \begin{center}\label{fig3}
   \epsfig{file=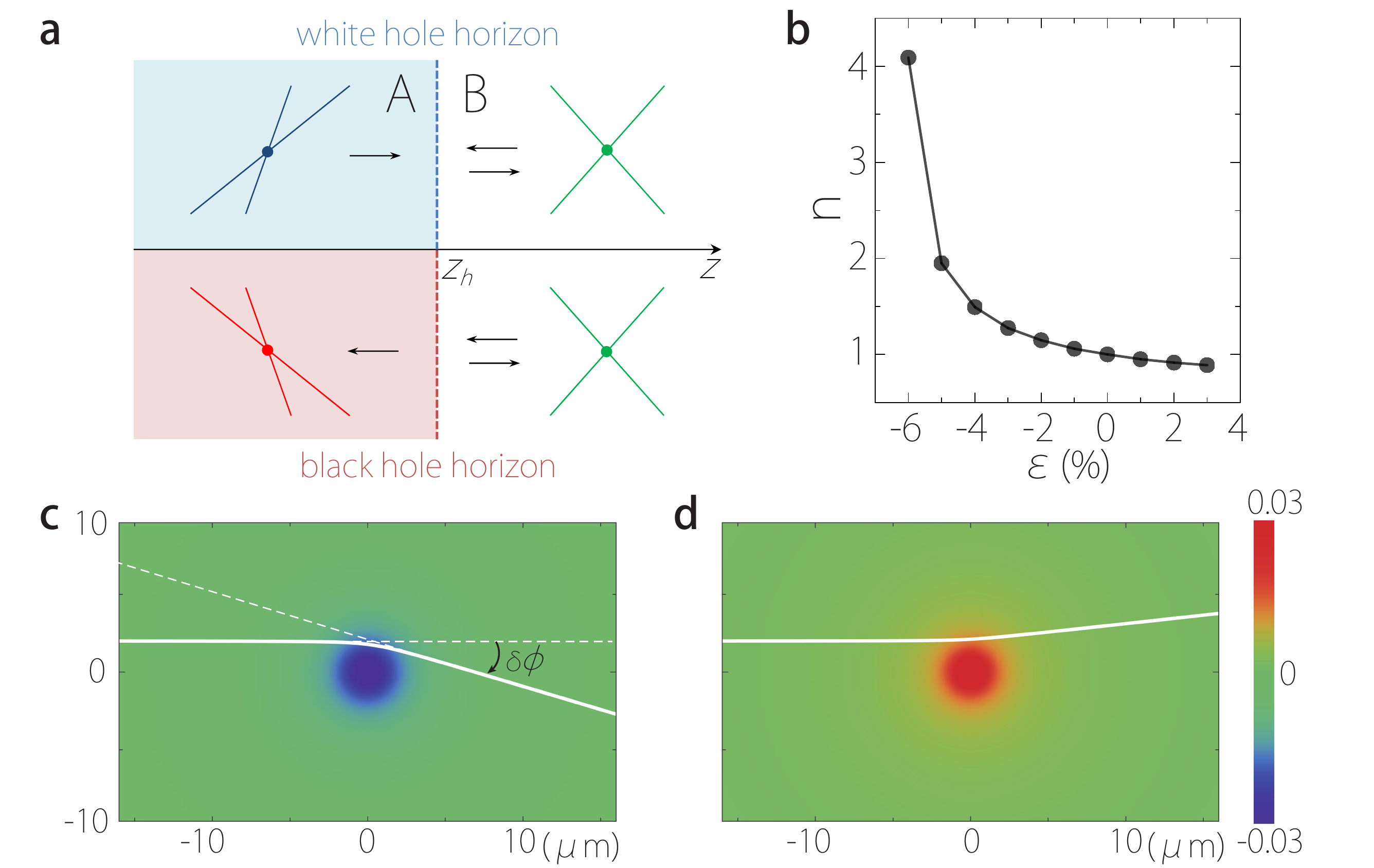,width=16cm}
  \end{center}
  \caption{\textbf{Astrophysical analogues in strained topological semimetal.} (\textbf{a}) Schematic figure showing (up) a white-hole horizon and (bottom) a black-hole horizon at $z=z_h$, corresponding to the Schwarzschild radius of metric (\ref{1D}). The arrows indicate the quasiparticle propagation directions in each region. (\textbf{b}) Effective refractive index $n$ for quasiparticle propagation versus strain. (\textbf{c,d}) Analogue of gravitational lensing effect. The white line indicates a quasiparticle trajectory (geodesic) in the $xy$-plane. $\delta\phi$ indicates the deflection angle. The colormap shows the strain profile. In (\textbf{b-d}), parameters of Na$_3$Bi are used. }
\end{figure}

\newpage
\begin{figure}
  \begin{center}\label{fig4}
   \epsfig{file=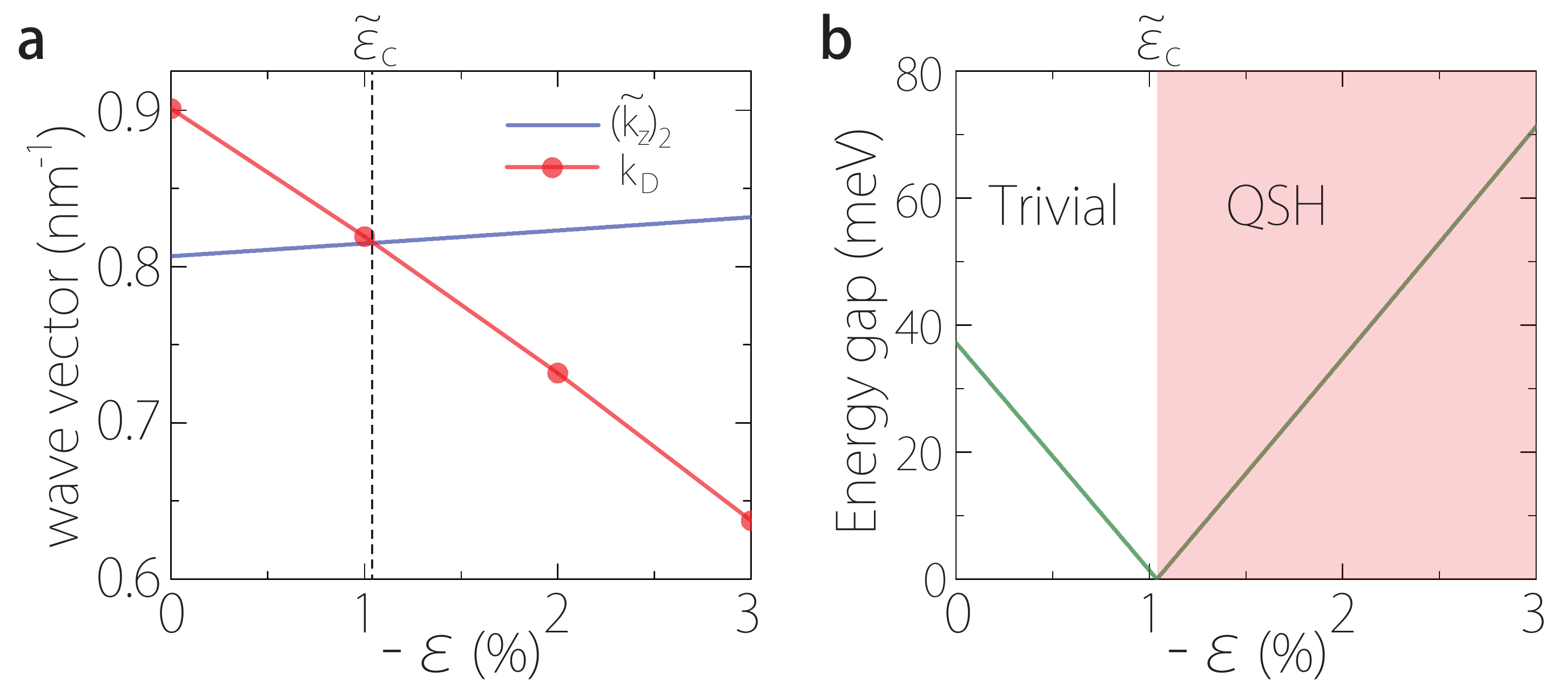,width=16cm}
  \end{center}
  \caption{\textbf{Topological phase transition in thin film.} (\textbf{a}) The quantum well effective wave-vector $(\tilde{k}_z)_2$ (for the second quantum well subband) and $k_D$ versus strain. Their crossing-point at critical strain $\tilde{\varepsilon}_c$ marks a topological phase transition between a trivial insulating and a QSH insulating phases. The corresponding subband gap closes at $\tilde{\varepsilon}_c$, as shown in (\textbf{b}). Here we take a Na$_3$Bi thin film with 16-layer thickness ($\approx 78$ \AA).}
\end{figure}
\newpage
\begin{figure}
  \begin{center}\label{fig5}
   \epsfig{file=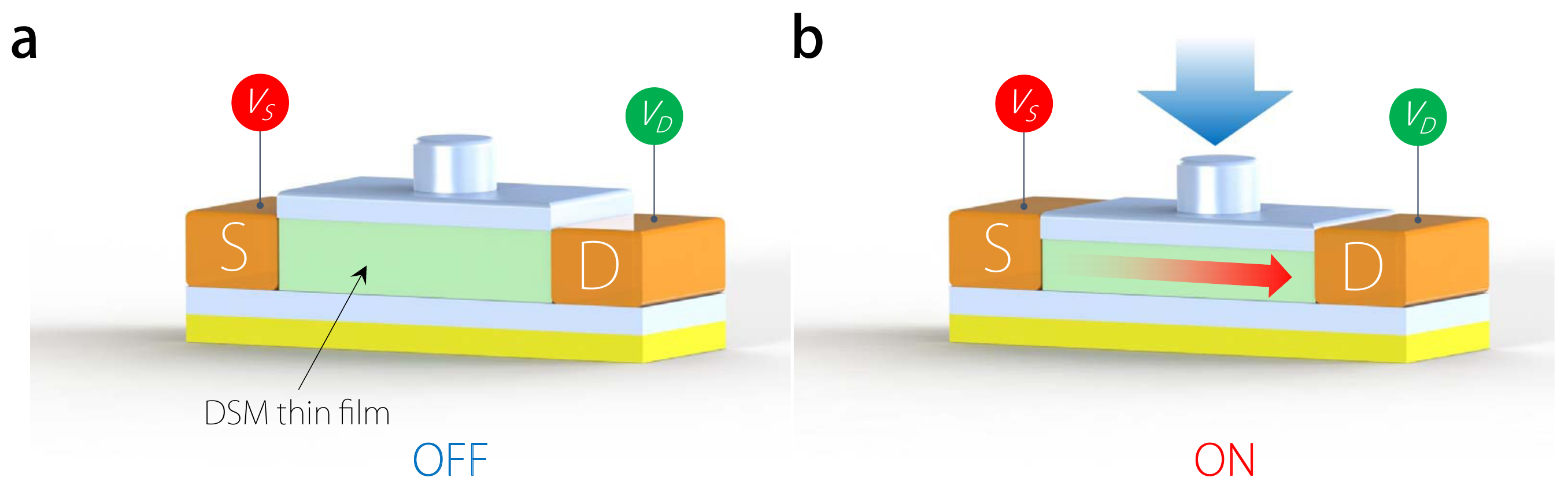,width=16cm}
  \end{center}
  \caption{\textbf{Piezo-topological transistor.} Schematic of a piezo-topological transistor using DSM thin film as the channel material. (\textbf{a}) The thin film is initially in the trivial insulating phase, so the transistor is in off state. (\textbf{b}) By applying strain to the channel region, the thin film is driven to the QSH phase with topological edge channels for conducting current. Hence the transistor is turned on. The topological channels are robust and have low heat dissipation.}
\end{figure}

\newpage
\begin{figure}
  \begin{center}\label{fig6}
   \epsfig{file=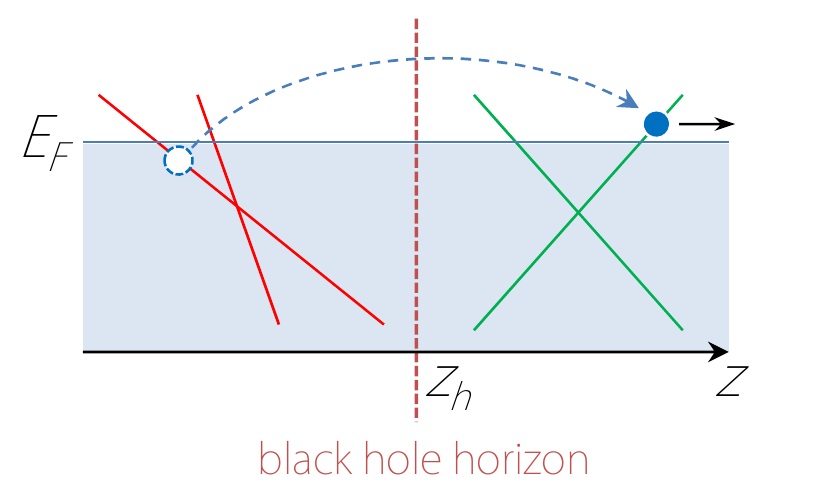,width=8cm}
  \end{center}
  \caption{\textbf{Analogue of Hawking radiation.} Illustration of the analogue of Hawking radiation in a topological semimetal. A tunneling process due to quantum fluctuation near an artificial black-hole horizon leads to the creation of quasipartile outside the Horizon and a quasihole inside the horizon. The quasiparticle escapes from the horizon, resembling the Hawking radiation. Here $E_F$ indicates the Fermi energy. }
\end{figure}

\end{document}